\documentclass[aps,amsfonts,pra,superscriptaddress,notitlepage,showpacs, twocolumn]{revtex4}

\usepackage{subfigure}
\usepackage{textcomp}
\usepackage{graphicx}
\usepackage{amssymb}
\usepackage{amsmath}
\usepackage{bm}
\usepackage{amsmath, amsthm, amssymb}
\usepackage{color}
\usepackage[colorlinks=true,linkcolor=blue,urlcolor=blue,citecolor=blue]{hyperref}

\def\beq{\begin{equation}}
\def\eeq{\end{equation}}
\def\bea{\begin{eqnarray}}
\def\eea{\end{eqnarray}}

\begin{document}

\title{Low-frequency conductivity in many-body localized systems}

\author{Sarang Gopalakrishnan}%
\affiliation{Department of Physics, Harvard University, Cambridge, MA 02138, USA}%
\author{Markus M\"uller}%
\affiliation{The Abdus Salam International Center for Theoretical Physics, Strada Costiera 11, 34151 Trieste, Italy}%
\author{Vedika Khemani}%
\affiliation{Department of Physics, Princeton University, Princeton NJ 08544, USA}%
\affiliation{Max-Planck-Institut f\"{u}r Physik komplexer Systeme, 01187 Dresden, Germany}%
\author{Michael Knap}%
\affiliation{Department of Physics, Harvard University, Cambridge, MA 02138, USA}%
\affiliation{ITAMP, Harvard-Smithsonian Center for Astrophysics, Cambridge, MA 02138, USA}%
\author{Eugene Demler}%
\affiliation{Department of Physics, Harvard University, Cambridge, MA 02138, USA}%
\author{David A. Huse}%
\affiliation{Department of Physics, Princeton University, Princeton NJ 08544, USA}%

\begin{abstract}
We argue that the a.c. conductivity $\sigma(\omega)$ in the many-body localized phase is a power law of frequency $\omega$ at low frequency: specifically,
$\sigma(\omega) \sim \omega^\alpha$ with the exponent $\alpha$ approaching 1 at the phase transition to the thermal phase, and asymptoting to 2 deep in the
localized phase.  We identify two separate mechanisms giving rise to this power law: deep in the localized phase, the conductivity is dominated by rare
resonant pairs of \emph{configurations}; close to the transition, the dominant contributions are rare \emph{regions} that are locally critical or in the thermal phase.
We present numerical evidence supporting these claims, and discuss how these 
power laws can also be seen through polarization-decay measurements in ultracold atomic systems.

\end{abstract}

\maketitle

\section{Introduction}

Isolated interacting quantum systems can undergo a dynamical phase transition---termed the many-body localization (MBL) phase transition---between a ``thermal'' phase in which the system comes to thermal equilibrium from generic initial conditions and a ``localized'' (or MBL) phase in which it does not~\cite{pwa, fleishman1980, gmp, baa, oh, znidaric, ph, nh}. Instead, an isolated system in the MBL phase is a ``quantum memory'', retaining some local memory of its local initial
conditions at arbitrarily late times~\cite{va, spa, hno, bvav, ckls,hnops, bn, rms, kbp, skg}.
The existence of the MBL phase can be proved with minimal assumptions \cite{jzi}; many of its properties are phenomenologically understood  \cite{spa,hno,rms},
and can be explored using the strong-randomness renormalization group~\cite{va, prado, ppv, VHA}.  While the eigenstate properties of MBL systems are
similar to those of noninteracting Anderson insulators, there are important differences in the dynamics, such as the logarithmic spreading 
of entanglement in the MBL phase~\cite{znidaric, bpm, va, spa2, hno, skg, kns}.

\begin{figure} [b]
\begin{center}
\includegraphics{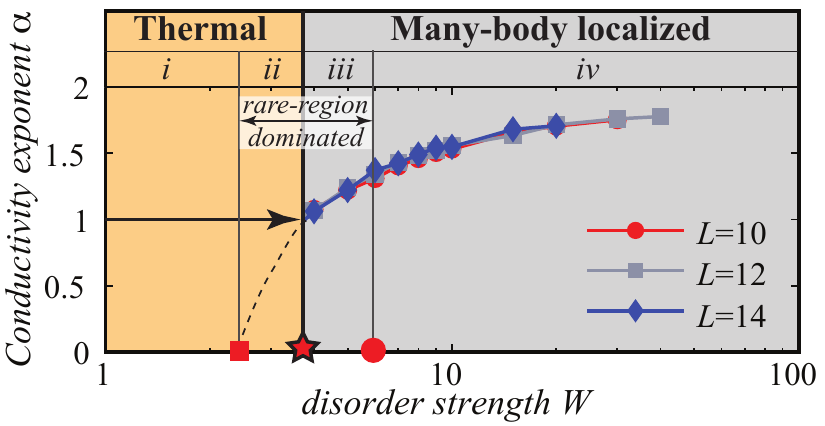}
\caption{A.C. conductivity exponent $\alpha$ of a disordered one-dimensional system across the many-body localization transition, showing four regimes:
(i)~the diffusive thermal phase; (ii)~the \emph{subdiffusive} thermal phase~\cite{agkmd}, which exists only in one dimension; and (iii, iv)~the MBL phase.
The MBL phase is divided into an ``MBL-Griffiths'' regime~(iii) in which low-frequency response is dominated by rare critical or thermal regions,
and an ``MBL-Mott'' regime~(iv) in which it is dominated by pairs of resonant configurations. Our main predictions are that the exponent
$\alpha \rightarrow 1$ [i.e., $\sigma \sim \omega$] as the MBL transition is approached from the localized side, and that $1 \leq \alpha < 2$ throughout the MBL phase.
These are consistent with numerical simulations of a nearest-neighbor random-field XXZ chain (shown in the plot).  In the thermal phase, finite-size effects are strong, and a more careful analysis~\cite{agkmd} is needed to extract the conductivity exponent.}
\label{summaryfig}
\end{center}
\end{figure}

In this work we show that the MBL phase also differs sharply from noninteracting localized systems in its low-frequency response.
We focus on a.c. conductivity in the MBL phase, for concreteness and to make contact with previous literature on solid-state systems
(e.g., Mott's law~\cite{mott1968}); however, as we argue below, our discussion directly extends to relaxation dynamics, which is more easily accessible in experimental studies using ultracold atoms~\cite{bloch2015}, polar molecules~\cite{junye}, nitrogen-vacancy centers~\cite{nv}, and other forms of ``synthetic'' matter. We identify two physical mechanisms
 underlying the slow response:
(a)~the presence of resonant pairs of charge or spin \emph{configurations}, connected by slow 
many-body rearrangements; and (b)~the presence within an MBL system of rare thermalizing regions, or ``inclusions'', that act as local heat baths for their surroundings. These mechanisms are absent in noninteracting systems: thus, the differences in transport between single-particle and many-body localization can be traced to the much larger \emph{connectivity} of the many-body Hilbert space. The two mechanisms we discuss involve local dissipative transport of the conserved densities (e.g., particle number),
and are thus distinct from the ``pure dephasing'' processes that cause the slow growth of entanglement within the MBL phase~\cite{hno, spa2}.

Our results for the a.c. conductivity are as follows. Whereas in noninteracting systems the a.c. conductivity
$\sigma(\omega) \sim \omega^2 \log^{d+1} \omega$ (Mott's law~\cite{mott1968} in $d$ dimensions), in the interacting MBL phase at high temperature the
conductivity goes as $\sigma(\omega) \sim \omega^\alpha$, where $\alpha$ is an exponent that varies continuously throughout the MBL phase,
ranging from $\alpha = 1$ at the MBL transition to $\alpha \rightarrow 2$ deep in the MBL phase. The exponent $\alpha$ has two regimes of behavior,
corresponding to the two mechanisms described above. Deep in the MBL phase, the conductivity is dominated, as in Mott's law, by 
resonant transitions between localized configurations.  It is enhanced relative to noninteracting localization because more such resonances are possible:
in addition to single-particle hopping, a MBL state can undergo multiple-particle rearrangements. We term this regime the ``MBL-Mott'' regime,
and argue that response in this regime is dominated by rare regions that are still localized but with anomalously large localization length.
Close to the transition, the conductivity is dominated by rare thermalizing or critical regions and their surroundings; we call this the ``MBL-Griffiths'' regime.
As the system approaches the transition from the insulating side, thermalizing inclusions proliferate; we show that this leads to the conductivity
exponent $\alpha \rightarrow 1$.

In two or more dimensions, the exponent $\alpha = 0$ throughout the thermal phase (i.e., there is presumably a nonzero d.c. conductivity).  In one dimension,
however, a subdiffusive phase with a continuously varying conductivity exponent $0 < \alpha < 1$ exists on the thermal side of the MBL
transition~\cite{blcr, agkmd, VHA}. Remarkably, therefore, the a.c. conductivity exponent $\alpha$ in one dimension is continuous and apparently smooth across the MBL transition,
approaching the critical behavior $\sigma(\omega) \sim \omega$ from both sides (Fig.~\ref{summaryfig}).

This paper is structured as follows. In Sec.~\ref{sec:assumptions} we list our assumptions. In Secs.~\ref{sec:mott} and~\ref{sec:griffiths} we introduce the MBL-Mott and MBL-Griffiths phases respectively; then in Sec.~\ref{sec:mottgriffiths} we discuss the transition between these phases. In Sec.~\ref{sec:numerics} we describe the numerical methods used to compare our theoretical predictions with data on random-field Heisenberg spin chains (details of the numerical analysis are given in Appendix~\ref{numericsdetails}). Sec.~\ref{sec:expt} connects the a.c. conductivity to the relaxation dynamics measured in ultracold atomic experiments. Finally, Sec.~\ref{sec:discussion} summarizes our results and comments on their scope.

\section{Assumptions}\label{sec:assumptions}

We work with a generic disordered lattice Hamiltonian having a conserved density (e.g., a particle number, or a particular projection of spin).  The current associated with this charge is denoted $\mathbf{j}$.  The a.c. conductivity tensor $\sigma$ in the $T \rightarrow \infty$ limit is then given by the Kubo formula:
\beq\label{kubo}
T\sigma_{\alpha \beta}(\omega) = \frac{1}{ZN}\sum_{mn} \langle m | j_\alpha | n \rangle \langle n | j_\beta | m \rangle \delta(\omega - \omega_{mn})~.
\eeq
Here $T$ is the temperature, $N$ is the number of sites, $Z$ is the partition function which in this infinite $T$ limit is the dimension of the many-body state space; the indices $m, n$ run over all $Z$ many-body eigenstates; and the current $j_\alpha$ is the sum over local currents, viz.
$j_\alpha \equiv \sum\nolimits_i j_{i, \alpha}$.  We shall only be concerned with the diagonal elements $\sigma_{\alpha\alpha}$,
so henceforth we shall drop the index $\alpha$.
Our arguments should also apply to the frequency-dependence of the a.c. thermal conductivity, e.g., in systems where the only conserved quantity is the energy.

When we consider the MBL phase, we specialize here to the case where all many-body eigenstates are localized, so we can discuss in terms of the localized conserved operators.  However, the results we obtain should also apply to the MBL phase in systems with a many-body mobility edge.  In the latter case, when we discuss `rare regions' they are not only rare local disorder configurations in the system's Hamiltonian, but also rare local configurations of the state that put it locally closer to, at, or across the mobility edge.

Our considerations here are at the level of linear response theory: i.e., we assume throughout that the drive is sufficiently weak and is present for a sufficiently short time that linear response applies. It has recently been shown \cite{kns} by one of the present authors that localized systems subject to a fixed-amplitude drive display a highly non-local response at low enough frequencies. Further, an MBL system subject to a finite-frequency drive for a long enough duration will eventually leave the linear response regime and enter instead a Floquet MBL steady state or even thermalize due to the ac drive~\cite{nh, kns, floquetMBL}. In the present work we are not concerned with these regimes.

\section{MBL-Mott regime}\label{sec:mott}

\subsection{Many-body ``Mott'' conductivity}

We begin by considering the generic behavior deep in the MBL phase.   As discussed earlier, we specialize to the regime where all eigenstates of the system are localized.
In this regime, the system Hamiltonian admits a representation in terms of effective spin-1/2 degrees of freedom labeled $\tau_k$, which are frequently referred to as local integrals of motion or ``l-bits''~\cite{spa,hno,rms}: in terms of these, $H = \sum_i h_i \tau^z_i + \sum_{i,j} J_{ij} \tau^z_i \tau^z_j + \ldots$.
Eigenstates of $H$ are also simultaneously eigenstates of all the $\tau^z_k$. These effective $\tau$ spins are related to the microscopic degrees of freedom
(which need not be spin-1/2) by a unitary transformation that is local up to exponentially small tails.
In terms of the effective $\tau$ spins, the current operator can be expressed as
\beq
j = \sum_{\alpha, k} K^{(1)}_{\alpha, k} \tau^\alpha_k + \sum_{\alpha, \beta, k, l} K^{(2)}_{\alpha\beta, kl} \tau^\alpha_k \tau^\beta_l + \ldots~,
\eeq
where $\alpha = x,y,z$; $\tau^\alpha$ is the appropriate Pauli matrix; and $k, l$ run over effective spins. The coefficients $K^{(n)}$ for $n\geq 2$ fall off
exponentially with the distance between the farthest effective spins in that term.  Stability of the MBL phase further requires them to fall off exponentially
with the order $n$~\cite{gn}.
Note that for a single-particle (noninteracting) Anderson insulator the $\tau^\alpha$ operators are the creation, annihilation and number operators of the localized single-particle states, so the coefficients $K^{(n)}$ are zero for $n > 2$, i.e., the current operator only contains single-particle hops and no multiple-particle rearrangements.

We now briefly review Mott's argument~\cite{mott1968} for the a.c. conductivity in \emph{noninteracting} localized systems at temperature $T>\omega$.
For this noninteracting case, the transitions contributing to $\sigma(\omega)$ at low frequency involve rare pairs of resonant \emph{sites} that hybridize to form pairs of nearly-degenerate eigenstates (i.e., symmetric and antisymmetric combinations of the wavefunctions centered at the two sites) with small energy splitting $\omega$.  Short-distance resonances, while common, typically have large splittings because of local level repulsion; these local processes give only a subdominant contribution to the conductivity in the low-frequency limit~\cite{footnote}.
To find the resonant pairs of sites with energy splitting $\omega$ that dominate in $\sigma(\omega),$ one has to go a distance $r_\omega$ determined by the condition $W \exp(-r_\omega) = \omega$, where $W$ is a microscopic bandwidth.  (We are assuming the single-particle localization length is of order
one lattice spacing and do not include factors of it.)  The number of such pairs is $\sim r_\omega^{d - 1}$.  The typical current matrix element between such pairs of eigenstates is $\omega r_\omega \simeq \omega \log (W/\omega)$ (because they involve moving a unit charge a distance $r_\omega$ at a rate $\omega$). Putting these pieces together, we recover the Mott result:
\beq
\sigma(\omega) \sim \omega^2 \log^{d + 1}(W/\omega)\,.
\eeq
Note that the contribution from more distant pairs is weaker, because the current matrix element falls off as $\exp(-R)$ whereas the phase space only grows as $R^{d - 1}$.

This argument is fundamentally altered by many-body processes for the \emph{interacting} MBL phase.
Here the conductivity includes not only hopping resonances between pairs of \emph{sites} but also many-body resonances between pairs of \emph{configurations};
hence the ``phase space'' factor is strongly enhanced. This argument is fundamentally altered by many-body processes for the \emph{interacting} MBL phase.
Here the conductivity includes not only hopping resonances between pairs of \emph{sites} but also many-body resonances between pairs of \emph{configurations}.
Hence the ``phase space'' factor is strongly enhanced. We now argue that this enhanced phase space factor grows exponentially in the number of effective spins flipped.

The many-body resonances that dominate the low-frequency dynamics flip $n$ effective spins,
with those spins typically having random spacings of order the localization length or less so that they do interact with each other.  For $d>1$ this set
of spins will in general have a fractal geometry.  Let $\gamma$ collectively denote all the relevant parameters (shape, typical interparticle spacing, etc.) specifying the ensemble of possibly resonant `clusters' of
flipped spins. (Given a cluster, in other words, one can characterize it through its parameters $\gamma$; different resonant $n$-spin clusters with the same $\gamma$ will have the same hybridization strength.)  

The typical current matrix element for a resonance with parameters $\gamma$ that flips $n$ spins is $\sim W\exp(-n/\zeta(\gamma))$.
Here $\zeta(\gamma)$ is a dimensionless quantity that depends on $\gamma$ and varies continuously in the MBL phase; $\zeta(\gamma)$ remains finite
at the MBL transition and decreases as one moves deeper into the localized phase.  $\zeta(\gamma)$ is larger for resonances having more closely spaced and thus
more strongly interacting spins.
Let us fix these parameters $\gamma$. Then, analogous to the single-particle case, the frequency $\omega$ picks out an ``optimal'' $n$ such that
\beq\label{mbmott}
W \exp(-n/\zeta(\gamma)) = \omega \Rightarrow n = \zeta(\gamma) \log(W/\omega).
\eeq
The number of possible resonances (in the ensemble parameterized by $\gamma$) that flip $n$ effective spins in the immediate vicinity of one particular real-space location is
exponential in $n$, while the frequency bandwidth of such rearrangements is linear in $n$.  Thus, to leading order the density of states of possible
resonances at order $n$ grows exponentially with $n$.
Specifically, it grows as $\sim e^{s(\gamma) n}$, where $s(\gamma)$ is the configurational entropy per flipped spin of the possibly resonant clusters in the
ensemble $\gamma$.  This is the entropy of all the possible choices of the $n$ spins flipped by the resonance.
Using this and Eq.~\eqref{mbmott}, the density of states of resonant configurations from ensemble $\gamma$ at frequency $\omega$ grows as a power law, $\omega^{-\phi}$,
where $\phi = s(\gamma)\zeta(\gamma)$, in contrast with the logarithmic growth in the noninteracting case.  The dominant resonances at low frequency flip many 
spins and have their properties $\gamma$ chosen so that the product
$s(\gamma)\zeta(\gamma)$ is maximized~\cite{fnstable}. 

We now assume that we have maximized this product $\phi$, and complete our estimate of the MBL-``Mott'' conductivity. The current matrix elements remain $\sim\omega$, up to logarithmic factors.  Putting this together with the phase space factor $\omega^{-\phi}$, the conductivity goes as
\beq
\sigma(\omega) \sim \omega^{2 - \phi}
\eeq
at low frequency. 

For the MBL phase to be stable,
we need that a typical eigenstate is, at a typical real-space location,
not involved in many resonances.
From the discussion above, the typical accessible phase space for final states with a matrix element of $\omega$ goes as $\sim\omega^{-\phi}$; thus the typical level spacing for these goes as $\sim\omega^\phi$.  In order that long-range resonances remain rare and do \emph{not} destabilize the MBL phase, the matrix element must vanish faster than the typical level spacing in the long-distance, small $\omega$ limit.  Thus $0<\phi<1$ (and thus $2>\alpha>1$) in the MBL phase, with $\phi$ increasing (thus $\alpha$ decreasing) as the phase transition to the thermal state is approached.

\subsection{Rare-region Mott resonances}

In the above discussion, we argued that the low-frequency conductivity in the MBL phase is dominated by rare many-spin resonances, and goes as $\omega^{2 - s \zeta}$, where $s$ and $\zeta$ are properties of the MBL phase (optimized over families of resonances parameterized by $\gamma$). However, in a disordered system, $\zeta$ is itself a random variable, so there will be atypical clusters in which (for example) the random fields are small and therefore the system is locally closer to the delocalization transition. (We focus on $\zeta$ but the same argument can be applied to any other parameter.) In such segments, $\zeta$ will take a local value $\zeta_{\mathrm{loc.}}$ that deviates from its typical value $\bar{\zeta}$, and the matrix element for resonances involving $n$ spins will be atypically large. 

These rare local ``regions'' occur with a probability $\sim\exp[-r f(\zeta_{\mathrm{loc.}})]$, where $f(\zeta_{\mathrm{loc.}})$ is a nonnegative ``rate
function''~\cite{dembo} that vanishes quadratically at $\zeta_{\mathrm{loc.}} = \bar{\zeta}$.  By the above arguments the contribution of such a rare local resonance to the ac conductivity will be $\sim\omega^2$, while the number of spins flipped by the rare resonance is $n \approx \zeta_{\mathrm{loc.}} \log (W/\omega)$ (which sets a minimum ``size'' for the rare region). Therefore the density of such rare resonances will be $\sim \omega^{(\zeta_{\mathrm{loc.}}f(\zeta_{\mathrm{loc.}}) - \zeta_{\mathrm{loc.}}s)}$.  Because the initial `gain' in conductivity by going to these rare local resonances is linear in $(\zeta_{\mathrm{loc.}}-\bar{\zeta})$ while the probability `cost' is only quadratic in the deviations from typical, the generic situation in the MBL phase is that the low-$\omega$ conductivity is dominated by rare many-body resonances in rare regions that are locally atypically close to the delocalization transition (i.e., have an atypically large $\zeta_{\mathrm{loc.}}$).

When the system is deep in the MBL phase, the dominant contributions to the low frequency conductivity are from resonant clusters in regions that are themselves
in the localized phase; we call this regime the ``MBL-Mott'' regime. In the low frequency limit in this regime, each resonant cluster is large compared to its local value of the localization length.  As the transition to the thermal phase is approached, at some point before
reaching the transition these dominant rare clusters become instead locally critical or thermal quantum Griffiths regions.
We now turn to such rare-region Griffiths effects, and show that they give rise to a conductivity exponent that approaches $\alpha = 1$ at the critical point.

\section{MBL-Griffiths regime}\label{sec:griffiths}

\begin{figure}[t]
\begin{center}
\includegraphics{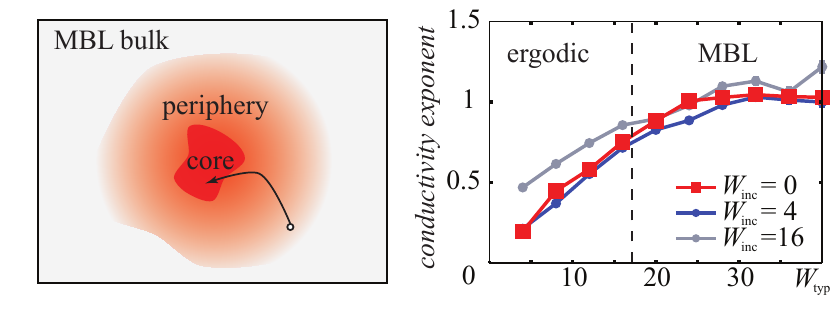}
\caption{Left: schematic of the structure of a thermal inclusion in the MBL phase, showing its core (the microscopically rare thermal region), the periphery (the typical surroundings that get strongly entangled with the rare region), and the typical MBL surroundings where the core does not flip the local effective spins.
Low-frequency transport occurs through transport between the periphery and the core. Right: numerically computed conductivity exponent $\alpha$
of an XXZ chain of size $L=12$ with integrability-breaking next-to-nearest neighbor exchange coupling that contains a thermal or critical inclusion
(i.e., a region that is locally thermal or critical) of four sites.  $W_\text{typ}$ is the typical value of disorder, 
and three different disorder values for the inclusion $W_\text{inc}$ are shown.  When the typical system is in the MBL phase ($W_\text{typ}>8$), the conductivity exponent saturates to near one, consistent with the discussion in the main text.}
\label{inclusions}
\end{center}
\end{figure}

We shall eventually be concerned with both thermal and critical rare regions, but to set up our discussion we begin by considering an inclusion that is locally
deep in the thermal phase, embedded in a typical insulating environment.  This thermal inclusion is of volume $V$, has a many-body level spacing $\Delta$ that
decreases exponentially with $V$ and a transport time (i.e., Thouless time) $t_{Th}$ that increases polynomially with $V$ (specifically, as $V^{2/d}$ for a compact
internally diffusive inclusion and with a larger power for a fractal or critical inclusion).
In general, $t_{Th} \Delta \ll 1$ for large thermal inclusions.  (We expect this also to be true for critical inclusions in $d > 1$.)  Moreover, each
inclusion thermalizes its immediate insulating surroundings.
Thus the inclusion consists of two parts:
first, the `core' of the inclusion which is the rare region that is locally thermalizing (or critical), and second, the typical insulating region surrounding this core, which gets strongly entangled with the core in the many-body eigenstates (we call this the `periphery').  A more thorough discussion of these inclusions is presented in Appendix A.

Now we consider the a.c. response of this inclusion, probed at a frequency $\omega \sim \Delta \ll 1/t_{Th}$.  Because the core relaxes rapidly compared with $\omega$,
it essentially adiabatically follows the applied electric field, and its response is reactive rather than dissipative~\cite{igb, si}.  Specifically, to leading order,
the core response goes as $\sigma(\omega) \sim \omega^2 t_{Th}$, which is subleading at low frequencies to the many-body Mott contribution.
%
Thus transport within thermal inclusion cores does not dominate the low-frequency conductivity.

However, the \emph{periphery} of an inclusion with core level spacing $\Delta$ does contribute strongly to its conductivity at frequencies down to $\Delta$, as we now argue.
This periphery consists of typical MBL regions that experience the core as a finite bath~\cite{ngh, jnb} to which they are coupled with matrix elements that fall off as
$\sim\exp(-R/\tilde\zeta)$ (where $R$ is the distance from the core and $\tilde\zeta$ is a decay length).  We can estimate the decay rate of a peripheral spin, using the Golden Rule, as
$\gamma(R) \sim W \exp(-2R/\tilde\zeta)$.  So long as $\gamma(R) \gg \Delta$, the Golden Rule is valid on these time scales and the core does indeed
act as a `bath' for these spins.  Far from the inclusion core, however, $\gamma(R) \ll \Delta$; a
spin at this distance resolves the discreteness of core levels and does not decay into them. The overall picture is as follows (Fig.~\ref{inclusions}): The core (with level spacing $\Delta$) is surrounded by `shells' of continuously decreasing $\gamma(R)$, with the outermost `active' shell having a decay rate $\gamma(R) \simeq \Delta$.  Beyond this distance the system is insensitive to the presence of the thermal core and remains fully localized.

This picture thus 
gives the behavior of the a.c. conductivity in the presence of a single such thermal inclusion.  When one probes the system at a frequency $\omega \geq \Delta$, the conductivity is dominated by the shell at radius $R_\omega$ such that $\gamma(R_\omega) = \omega$. Shells closer to the core relax faster, and their response to a probe oscillating at $\omega$ is mainly reactive; meanwhile, shells that are farther do not respond at all at $\omega$.  The conductivity of the dominant shell is proportional to its Golden-Rule decay rate, so this shell gives $\sigma \sim \gamma(R_\omega) \sim \omega$ (up to $\log\omega$ factors due to size, dipole matrix element, etc.).  The conductivity due to a single inclusion thus turns on at and above a frequency $\Delta$ and has the behavior $\sigma \sim \omega$ at intermediate frequencies. This reasoning extends directly to any inclusion whose core has an internal relaxation rate greater than its many-body level spacing, and is supported by numerical simulations (Fig.~\ref{inclusions}b) on inhomogeneous
systems, in which thermal or critical inclusions are put in by hand: the conductivity in the presence of an inclusion goes as $\sigma \sim \omega$ whenever the core is thermal or critical, as expected.

We now use the single-inclusion result to study the rare-region contribution to the a.c. conductivity of a generic MBL system, which contains some density of inclusions at all scales.  The cores of these inclusions must be thermal, but there does not appear to be any constraint on how thermal they are, so the most common such cores of a given size will be cores that are locally arbitrarily close to the critical point.  The
inclusion cores that dominate the conductivity at low $\omega$ are thus rare locally critical regions with level spacing $\omega$ and consequently of
volume $\sim \zeta_c\log(W/\omega)$; the probability of such cores is therefore $\sim p^{\zeta_c\log(W/\omega)}\sim\omega^g$, where $p$ is (heuristically)
the probability that a unit-volume region is locally critical. (One can define $p$ more precisely as follows:
the density of critical inclusions of volume $V$ decreases exponentially with $V$, as $p^V$.)
In the MBL phase $p<1$, and $p$ approaches one at the transition; thus the Griffiths exponent $g$, which is positive, approaches zero as the transition is approached.
Since each such inclusion contributes $\sim\omega$ (up to logarithmic corrections) to the conductivity, the resulting conductivity of the Griffiths insulator goes as $\sigma\sim\omega^{1 + g}$, where the Griffiths exponent $g$ goes to zero at the critical point and rises smoothly in the MBL phase.

We briefly comment on how these Griffiths arguments connect with those in the thermal phase~\cite{agkmd, VHA}. In the MBL phase, as discussed above, thermal
inclusions of large volume $V$ become exponentially rare in $V$.  On the thermal side, instead, it is \emph{localized} inclusions that become exponentially
rare at large scales. In one dimension, these rare localized inclusions act as transport bottlenecks, leading to a subdiffusive Griffiths phase~\cite{agkmd}.
In higher dimensions, however, rare insulating regions in the thermal phase cannot block transport,
and the d.c. conductivity is nonzero.

\section{Transition between MBL-Mott and MBL-Griffiths regimes}\label{sec:mottgriffiths}

The overall behavior of the low-frequency conductivity exponent $\alpha$ is shown in Fig.~\ref{summaryfig}:
everywhere in the MBL phase 
$1 < \alpha < 2$.  Near the MBL phase transition, critical and thermal inclusions proliferate and the
dominant mechanism is ``Griffiths''; far from the MBL transition, such inclusions are too rare, and the dominant contributions to conductivity are instead from rare
many-body Mott resonances within locally insulating regions (which are still less insulating than the \emph{typical} region).  These regimes transition into each
other as follows:  Within the ``MBL-Mott'' phase the dominant regions are still locally insulating; as one moves towards the transition, these dominant regions become
less insulating.  Eventually, before the MBL phase transition, the dominant regions become critical, and the system enters the ``MBL-Griffiths'' phase.  Throughout the MBL-Griffiths phase, the dominant rare regions remain critical, and only their prevalence changes: as the critical point is approached, these rare critical regions become more common, and eventually proliferate.  Thus the physics underlying the evolution of the conductivity exponent is qualitatively different in the two regimes, and we expect that the exponent is nonanalytic (though perhaps quite smooth) at the ``Mott-Griffiths'' transition between these two regimes. The location of the Mott-Griffiths transition line in Fig. 1 is schematic---determining the location of this transition within the MBL phase remains an interesting direction for future work.

\section{Numerical simulations}\label{sec:numerics}

\begin{figure}[t]
\begin{center}
\includegraphics[width = 0.45 \textwidth]{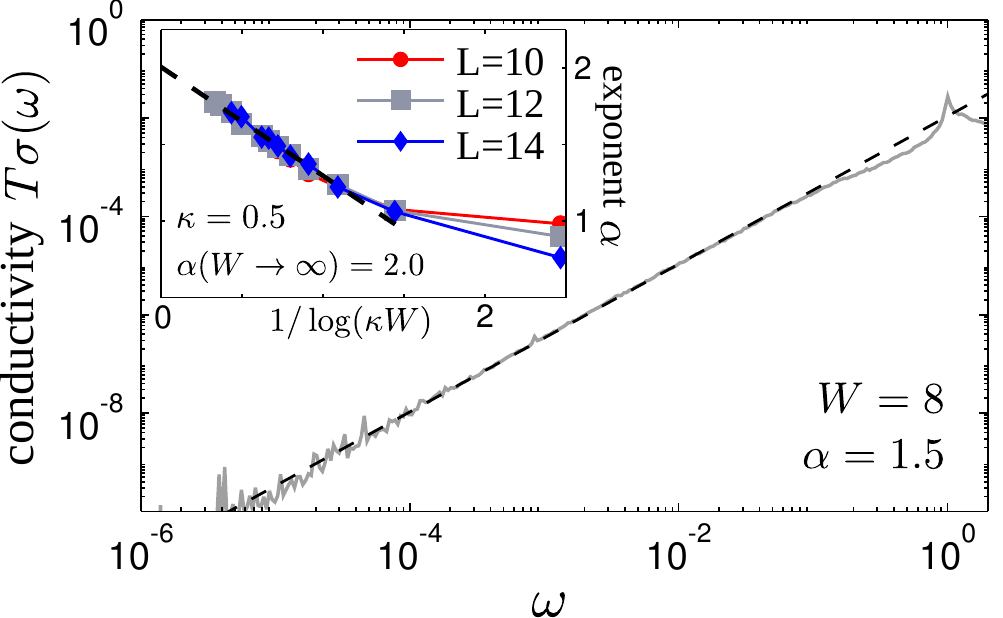}
\caption{Main panel: conductivity as a function of frequency in the random field XXZ chain for system size $L = 14$ and disorder bandwidth $W = 8$,
showing multiple decades of power-law behavior. Inset: behavior of the conductivity exponent deep in the MBL phase is consistent with the many-body Mott prediction
(see main text) that $2 - \alpha \sim \zeta$, combined with the perturbative estimate $\zeta \sim 1/\log(W/J)$.}
\label{conductivity-details}
\end{center}
\end{figure}

We have checked these expectations against simulations of the conductivity in the random-field XXZ chain, governed by the Hamiltonian
\begin{equation}
 \hat H = \frac{J}{2} \sum_{\langle ij \rangle} (\hat S_i^+ \hat S_{j}^- + \text{h.c.}) + J_z \sum_{i} \hat S_i^z \hat S_{i+1}^z  +  \sum_i h_i  \hat S_i^z\;,
 \label{eq:h}
\end{equation}
where $h_i$ is a local quenched random field drawn uniformly from $[-W,\,W]$, $J$ is the spin exchange energy scale, and $J_z$ the spin-spin coupling strength.  We measure energies in units of $J$; in all presented data, we also take $J_z=J=\hbar=1$.  We diagonalize the full Hamiltonian to calculate the conductivity $\sigma(\omega)$ at infinite temperature using Eq.~\eqref{kubo} by binning it on a logarithmically spaced frequency grid which typically ranges from $\omega = 10^{-6}$ to $2$.  In Figs. \ref{summaryfig} and \ref{conductivity-details} we have only nearest-neighbor exchange, while we also have second-neighbor exchange of strength $J'=1$ in Fig. \ref{inclusions}. The rationale for the latter choice is that, while next-nearest neighbor exchange interactions exacerbate finite-size effects, they also break integrability and make the disorder-free delocalized phase behave thermally, even for relatively small system sizes. Thus they are essential for correctly simulating small
thermal inclusions with $W_\text{inc}=0$.

The conductivity exponent $\alpha$, shown in Fig. \ref{summaryfig}, has been extracted from power-law fits to the low frequency response that hold over multiple decades, see Fig. \ref{conductivity-details} main panel for the example of $W=8$ which yields $\alpha\sim 1.5$. The numerical results further confirm that the conductivity exponent $\alpha = 1$ at the MBL transition and that it asymptotes to $\alpha \sim 2$ in the strong disorder limit, see inset of Fig. \ref{conductivity-details}. A more detailed discussion of some numerical issues and a comparison between noninteracting and
many-body insulators is given in Appendix B. 

\section{Experimental aspects}\label{sec:expt}

The predictions in our work concern the finite-time dynamical properties of MBL systems; thus, they are robust against weak coupling to an external bath, which is present in all physical systems~\cite{ngh}; so long as $\omega > \Gamma$, where $\Gamma$ is the bath-induced linewidth, the bath will not change these conductivity power laws.
Therefore our predictions for conductivity can be tested experimentally, both in solid-state systems~\cite{shahar2014} and in ultracold atomic systems~\cite{bloch2015}. In electronic systems, a.c. conductivity is straightforward to measure, but the long-range nature of the Coulomb interaction will modify several of our conclusions.

In principle one can also measure a.c. conductivity in ultracold atomic systems such as optical lattices by applying a periodically modulated tilt to the
entire lattice~\cite{higgs}.
However, in the current optical-lattice MBL experiments~\cite{bloch2015} it is more convenient to study relaxation in the time domain; we now show how our results
generalize to such experiments.  (Note: Ref~\cite{bloch2015} used a quasiperiodic potential, while we are discussing the case of a random potential.)
In general, these experiments involve creating a particular nonequilibrium density configuration and measuring the evolution of its ``contrast'' (i.e., the overlap between final and initial density deviations from thermal equilibrium).  In the MBL phase, this contrast (which we denote $C(t)$) approaches a nonzero saturated value $C_{\infty}$; we argue that it does so at long time $t$ as
\beq\label{bloch}
C(t) - C_{\infty} \sim t^{1 - \alpha},
\eeq
where $\alpha > 1$ is the a.c. conductivity exponent discussed here and plotted in Fig.~\ref{summaryfig}.  This result holds in both the MBL-Mott
and MBL-Griffiths regimes.  In the Mott regime, at time $t$, resonant pairs with splitting $\alt 1/t$ are still in their initial state and retain
their initial density deviation, whereas faster pairs oscillate and thus have ``forgotten'' their initial density deviation.  Counting all Mott pairs with splitting
$\alt \omega$, using the arguments above, we find that these go as $\omega^{1 - \phi} = \omega^{\alpha - 1}$, which gives Eq.~\eqref{bloch}.  Likewise, in the
Griffiths regime, the contribution at time $t$ is due to the peripheral spins of inclusions with core level spacing $\Delta \alt 1/t$.  The density of such
inclusions is $\sim \Delta^{\alpha - 1}$, which once again yields the result~\eqref{bloch}.  Note that this decay becomes very slow as the transition is approached:
the exponent $(1-\alpha)$ goes to zero at the transition.  These arguments apply to the long time behavior when $C(t)$ is near $C_{\infty}$; the earlier time regime
near the critical point when $C_{\infty}$ is small should be governed by the dynamical critical behavior.  Preliminary numerical simulations
on XXZ chains suggest that, deep in the MBL phase, this rare-region contribution might be difficult to detect in experiment, as its amplitude is small compared with
steady-state fluctuations of the contrast in finite systems.

\section{Discussion}\label{sec:discussion}

In this paper we have argued that the low frequency a.c. conductivity in the MBL phase goes as $\sigma(\omega) \sim \omega^\alpha$, with $1 \leq \alpha < 2$ throughout the phase, and $\alpha \rightarrow 1$ as the delocalization transition is approached.  Deep in the MBL phase, the dominant processes involve transitions between rare configurations, in rare regions that are localized but have an anomalously large localization length.  Near the transition, the dominant rare regions are locally thermal or critical instead.  The power-laws we expect on general grounds are consistent with those seen in numerical results for the random-field XXZ model.  We emphasize that the power laws we find in the optical conductivity are not related to those predicted for electron glasses~\cite{se}: we are considering high-temperature behavior (i.e., $\omega \ll T$) in models with short-range interactions, whereas those works consider low-temperature behavior (i.e., $\omega \gg T$) in models with Coulomb interactions.

We conclude with some comments on the scope of our results. As already discussed above, our analysis of a.c. transport directly extends to
relaxation dynamics. Moreover, our results here should also describe, e.g., thermal transport, in systems where the only conserved quantity is the energy.
However, our analysis of the conductivity relies on the fact that the conductivity is related to the spectral function of a \emph{current}
(i.e., a quantity associated with a globally conserved charge) and does not extend to generic spectral functions, such as those probed using optical
lattice modulation spectroscopy~\cite{higgs}.

Our analysis also depends on both the disorder correlations and the interactions being short-range.
Specifically, we assume that the effective interactions that mediate many-body resonances fall off exponentially with distance and with the number of effective spins involved.
Thus our conclusions are
modified in an essential way when the interactions instead fall off as a power law of distance; this case will be treated elsewhere.  It is not presently clear whether or not the stretched-exponential effective interactions that occur at putative critical points within the MBL phase~\cite{va, hnops, prado, ppv} substantially modify the above story.
Also, our analysis of near-transition behavior assumes that the delocalized phase is thermal, and thus may not apply to hypothesized transitions between an MBL phase and a nonthermal delocalized phase~\cite{dlaks, grover, cl}.

\section*{Acknowledgments}

We thank Dmitry Abanin, Ehud Altman, Boris Altshuler, Immanuel Bloch, Soonwon Choi, Michael Fogler, Igor Gornyi, Christian Gross, Lev Ioffe, Alexander Mirlin, Rahul Nandkishore, Vadim Oganesyan, Ulrich Schneider, and Norman Yao for useful discussions. We especially thank Kartiek Agarwal and Shivaji Sondhi for several enlightening conversations on closely related topics during previous collaborations. This research was supported by Harvard-MIT CUA, ARO MURI Fundamental Issues in Non-Equilibrium Dynamics, AFOSR MURI New Quantum Phases of Matter, ARO MURI Atomtronics, ARO MURI Quism program, the National Science Foundation Grant Nos. NSF PHY11-25915, NSF DMR-1308435 and NSF DMR-1311781, the Humboldt Foundation, the Max Planck Institute for Quantum Optics, the John Templeton Foundation, and the Austrian Science Fund (FWF) Project No. J 3361-N20.

\appendix

\section{Structure of a thermal inclusion}\label{inclusiondetails}

A thermal inclusion core (i.e., a large region with rare microscopic parameters) in the MBL phase acts as a local, discrete ``bath'' for the peripheral insulating material around it.  Thus, in the many-body eigenstates it is strongly entangled with the nearby (``peripheral'') typical regions.  Because the inclusion is finite, sufficiently far from it the MBL phase with area-law eigenstate entanglement re-establishes itself.  In this Appendix we discuss how this crossover takes place.

A naive estimate (which will turn out to be largely correct) is as follows:  If one ignores the discreteness of the bath levels, a Golden Rule estimate~\cite{ngh} suggests that the decay rate of a typical degree of freedom (which for convenience we shall call a spin) a distance $R$ from the inclusion core is $\sim \exp(-2R/\tilde\zeta)$ where the decay length $\tilde\zeta$ remains finite as the MBL transition is approached.  This rate must be compared with the many-body level spacing $\Delta \sim e^{-\tilde{s}V}$ of the inclusion core; when this putative decay rate is smaller than the many-body level spacing, the inclusion is actually unable to act as a bath~\cite{ngh} and no decay takes place.  Thus, an inclusion core of {\it volume} $V$ is surrounded by a thermal periphery of {\it linear size} $\sim V$.  The characteristic relaxation rate decays from its value at the center of the core (which is of order the bandwidth for a strongly thermal inclusion) to its value at the edge of the periphery, which is $\sim\Delta$.

However, this argument is evidently incomplete.  When the inclusion core thermalizes a spin, on sufficiently long time scales this additional spin is also ``thermal'', and thus naively might be thought able to act as a bath for other, more distant, spins.
If one iterates this reasoning, however, one arrives at an obviously incorrect result: an inclusion core of linear size $L$ thermalizes a region of linear size $\sim L^d$ around it; and the combined level spacing of this full thermalized region is now $\sim e^{-sL^{d^2}}$ ($s$ being the thermal entropy per spin), which naively allows it to thermalize yet further regions, and so on, until the entire MBL system is thermalized.  To avoid this conclusion, one must understand why these peripheral regions that are ``thermalized'' by the core cannot act as a bath for more distant insulating regions.

One can see this as follows:  Let us first remove all couplings that cross the boundary between the thermal inclusion core and the periphery.  Then the Hamiltonian of the now MBL peripheral region can be written in terms of l-bits, in terms of which it takes the fully diagonal form $H = \sum\nolimits_i h_i \tau^z_i + J_{ij} \tau^z_i \tau^z_j + \ldots$; the thermal inclusion core is, of course, described by a generic thermalizing Hamiltonian.  Now we reinstate the boundary couplings; these are local in terms of the physical spins, and thus generally consist of a strictly local physical operator $O$ on the MBL side of the boundary, coupled to an operator on the thermal side.  The operator $O$ involves l-bit flips at all distances, but contributions from distant l-bits are exponentially suppressed.  Because the intrinsic l-bit Hamiltonian is purely diagonal, an l-bit at a distance $l \gg \tilde\zeta$ from the boundary can only thermalize through its exponentially weak contribution to the operator $O$; in particular,
the nearer l-bits do not act as a bath, and the ability of an inclusion to thermalize its surroundings is determined by the size of its core.

\begin{widetext}

\section{Details of conductivity numerics}\label{numericsdetails}

In this section, we discuss some of the subtler issues involved in numerically extracting the conductivity exponents. We begin with a discussion of finite-size effects and boundary conditions. We then compare our many-body a.c. conductivity numerics with a study of single-particle (i.e., noninteracting) insulators at similar system sizes. We find that the size- and disorder-dependence of the many-body conductivity is consistent with theoretical expectations, and qualitatively different from that of the single-particle conductivity. Finally, we present data elucidating the nature of transitions contributing to the low-frequency conductivity.

\subsubsection{Boundary conditions and finite-size effects}

In numerics on finite systems, the conductivity exponents discussed here only occur at intermediate frequencies,
$\omega_L \ll \omega \ll J$, where $\omega_L \sim \exp(-L/\zeta)$ in the MBL phase [or $\exp(-L/\xi)$ in the single particle case] is a size-dependent low-frequency cutoff.
$\omega_L$ sets the scale for level repulsion between states or configurations that differ on length scales on the order of the system size.  The behavior below this frequency scale
depends on whether the boundary conditions are open or periodic.  In the case of open boundary conditions, the conductivity at the lowest frequencies goes as
$\sigma(\omega) \sim \omega^3$.  As in the Mott argument, two factors of $\omega$ are due to the current matrix element, which is constrained by the boundary conditions to vanish as $\omega$
at low frequencies.  The third factor is due to level repulsion in the Gaussian Orthogonal Ensemble~\cite{si}, and captures the phase space of pairs of states with these very small energy
differences.  For periodic boundary conditions, on the other hand, the conductivity in the finite-size-dominated regime scales as $\sigma(\omega) \sim \omega$.
In that case, there is still a factor of $\omega$ from level repulsion, but the current matrix elements do not vanish in the limit of low frequencies, because these
currents ``wrap around'' the system and hence do not build up large charge imbalance even when they are at very low frequency.
(The distinction between the two kinds of boundary condition can be intuitively understood by contrasting the behavior of (a)~a finite metallic grain embedded in an insulator and subject to spatially uniform a.c. electric field, and
(b)~a conducting ring with an a.c. magnetic flux through it.  The response of the former becomes essentially dielectric in the limit of low frequencies,
whereas that of the latter remains dissipative.)

\begin{figure*}[h]
\begin{center}
\includegraphics[width = 0.95\textwidth]{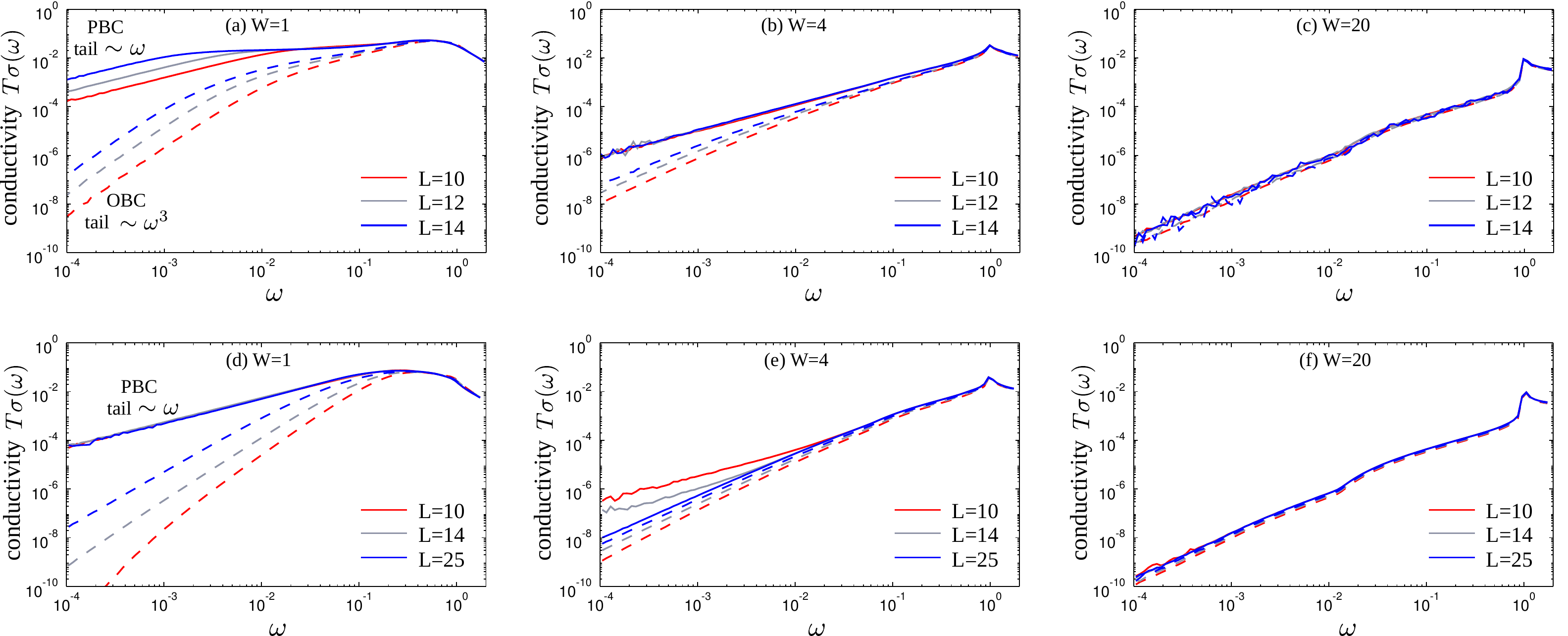}
\caption{(a)-(c)~Frequency-dependence of a.c. conductivity in the interacting system with periodic and open boundary conditions (solid and dashed lines respectively) for three values of disorder $W$, corresponding to the thermal phase (left), near-critical regime (center), and deeply localized phase (right). The lowest-frequency finite-size behavior in the thermal regime goes as $\sigma \sim \omega$ for periodic boundary conditions and $\sigma \sim \omega^3$ for open boundary conditions, as discussed in the text. (d)-(f)~Frequency dependence of a.c. conductivity in noninteracting systems, for disorder values corresponding to those in panels~(a)-(c). Again, solid lines represent periodic boundary conditions and dashed lines represent open boundary conditions.}
\label{MBL:boundaryconditions}
\end{center}
\end{figure*}

This finite-size dominated regime is clearly seen in numerical simulations on the many-body system in the thermal regime [Fig.~\ref{MBL:boundaryconditions}(a)].
Deep in the localized regime, $\sigma(\omega)$ is insensitive to boundary conditions in the frequency range we can access [Fig. \ref{MBL:boundaryconditions}(c)]; the finite size
effects presumably appear only at even lower frequency $\omega_L \sim \exp(-L/\zeta)$.
In the near-critical regime ($W = 4$), where $\alpha$ is near one, the finite-system behaviors for open and periodic boundary conditions are qualitatively different
[Fig.~\ref{MBL:boundaryconditions}(b)]:  For periodic boundary conditions, the low frequency regime due to the finite size effect is expected to also have
$\alpha=1$.  No finite size effect is apparent at $W=4$ for periodic boundary conditions, which is possibly a consequence of the finite size regime having
essentially the same scaling as the ``bulk'' regime.
On the other hand, for open boundary conditions the low frequency finite size regime will have an effective exponent $\alpha_\text{eff}=3$.  The beginnings of the crossover in to this regime are apparent in Fig.~\ref{MBL:boundaryconditions}(b), and we can see the results are converging with increasing $L$ towards the periodic boundary condition results.

The analogous results for noninteracting systems are shown in Fig. \ref{MBL:boundaryconditions}(d)-(f).  Deep in the localized phase at $W = 20$ the conductivity at these frequencies is dominated by short distance single-particle hops, so the interacting and noninteracting systems look similar and neither show finite-size effects.  At $W = 4$, on the other hand, we see clear differences.  The effective exponent $\alpha$ for the noninteracting system is well above one, so now the finite size effects are quite apparent for periodic boundary conditions.

\begin{figure*}[h]
\begin{center}
\includegraphics[width = 0.85\textwidth]{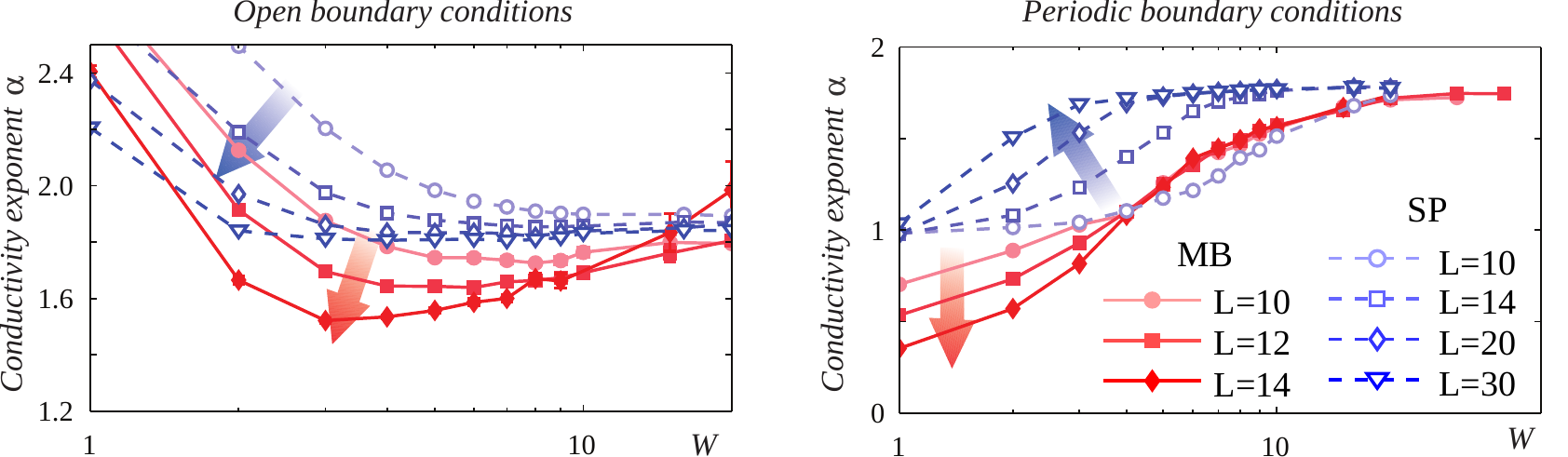}
\caption{Conductivity exponent for interacting spin chains (solid red lines) and the corresponding noninteracting chains (dashed blue lines) as a function of
disorder $W$, for various system sizes. The left panel shows data for open boundary conditions (OBCs); the right panel, for periodic boundary conditions (PBCs).
Exponents are extracted from the frequency regime $10^{-4} \leq \omega \leq 5 \times 10^{-3}$ [OBC] and $10^{-4} \leq \omega \leq J^2/(4W)$ [PBC]. Arrows indicate
the evolution of the exponent with increasing system size. For OBCs, the exponent crosses over from the finite-size value $\alpha = 3$ to the Mott value
$\alpha \simeq 2$ with increasing disorder. The crossover is nonmonotonic for interacting systems but monotonic for noninteracting systems, as discussed in the text. For PBCs,
noninteracting systems again exhibit a monotonic crossover from the finite-size exponent $\alpha = 1$ to the Mott exponent $\alpha = 2$. However, the
exponent for interacting systems drops well below the finite-size value $\alpha = 1$, in the regime where these systems are thermal.  Moreover,
finite-size effects on $\alpha$ seem negligible throughout the MBL phase.}
\label{comparison}
\end{center}
\end{figure*}
\begin{figure*}[h]
\begin{center}
\includegraphics[width = 0.95\textwidth]{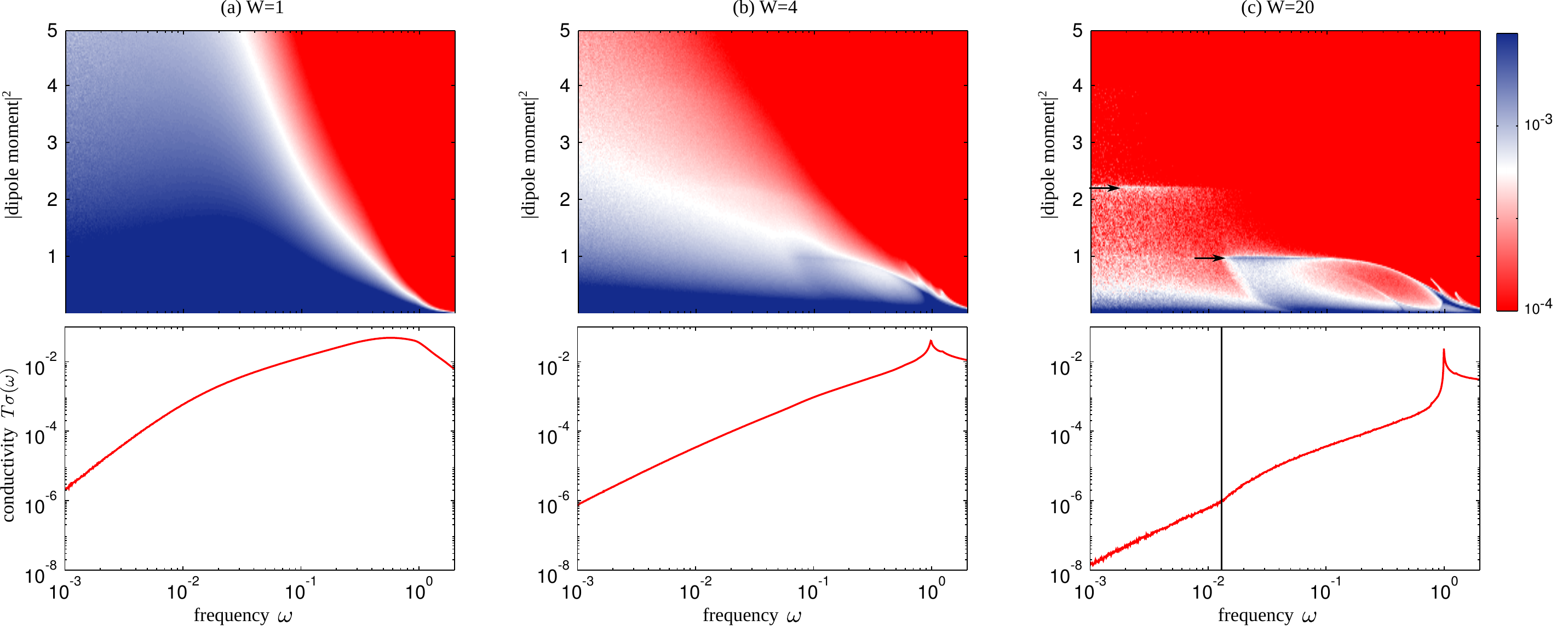}
\caption{Distribution of squares of dipole moment matrix elements (i.e., $| \langle n|\sum_j j S^z_j|m\rangle|^2$) contributing to the conductivity at each
frequency, for disorder corresponding to thermal phase (left), critical region (center), and MBL phase (right), all with open boundary conditions.
In (c)
the arrows mark the peaks corresponding to two-site and three-site hops in the MBL phase (see text); we expect the $n$-site peak to vanish at a frequency
$\sim 1/W^{n - 1}$, giving rise to a ``shoulder'' in the conductivity. Lower panel: frequency-dependence of the conductivity. %
 Note that the different regimes of behavior in $\sigma(\omega)$ can be matched with features in the dipole moment distribution.}
\label{dipoles}
\end{center}
\end{figure*}

These considerations can be sharpened by comparing the conductivity exponents extracted from the many-body (MB) interacting and single-particle (SP) noninteracting data, as shown in Fig.~\ref{comparison} (left).  We consider systems with open boundary conditions, and extract the conductivity exponent from a fixed frequency range that is much smaller than the scales $J^2/W, J^3/W^2$ associated with short-distance hops.  From our previous discussion, we expect that the limiting behavior for very small disorder is $\sigma \sim \omega^3$ (because of finite size effects), whereas that for large disorder is $\sigma \sim \omega^2$ (as finite size effects move to much lower frequencies and the Mott behavior is recovered).  In the single-particle case, one expects the exponent $\alpha$ to cross over smoothly from a disorder-independent value slightly below 2 -- on account of the logarithmic correction in Mott's law -- toward 3, and thus to increase monotonically as the disorder is decreased.  On the other hand, for
MBL, we expect that as the disorder is decreased two competing effects occur: the
exponent decreases towards 1 for the reasons discussed in the main text; on the other hand, it is also pulled up toward 3 by finite-size effects.  Thus, we expect it to exhibit a non-monotonic U-shaped disorder-dependence with a minimum near the MBL transition.  These expectations are borne out by the numerical data (Fig.~\ref{comparison}): the dip of the many-body exponent below 2 becomes \emph{stronger} for larger system sizes, approaching the value for periodic boundary conditions, and thus supporting the view that the true exponent for the many-body case is disorder-dependent and dips well below 2.

As a final point of comparison, Fig.~\ref{comparison} (right) shows the conductivity exponents for several different system sizes and as a function of disorder, extracted from the data with periodic boundary conditions.  For these exponents, we do the fit over several decades of data between $\omega_L$ and the microscopic scales $J^2/W$.  Thus, these exponents can be directly compared to the analogous MB ones plotted in Figs. ~\ref{summaryfig} and \ref{conductivity-details}.  We see that for any given system size, the SP data look qualitatively similar to the MB data, showing a monotonic increase from the finite-size dominated exponent $\alpha \sim 1$ at low disorder (note that $\alpha \sim 1$ at the MB transition due to a completely different physical mechanism) to an exponent approaching 2 at larger disorders.  However, unlike the MB system, the SP effective exponents are strongly finite-size dependent and approach a constant $W$-independent value on increasing system size.  On the other hand, the MB exponents show no
system size dependence in the
localized phase and converge to $W$-dependent values (significantly less than 2 for moderate disorder), further supporting our claim that the MB exponents are not finite-size effects.

\subsubsection{Nature of transitions contributing to conductivity}

We conclude with some details about the structure of the eigenstates contributing to the conductivity. The conductivity always exhibits a sharp feature near
$\omega = 1$, which is due to nearest neighbor resonances.  For strong disorder (Fig. \ref{MBL:boundaryconditions}c), the conductivity also develops a noticeable ``shoulder'' at a frequency
$\omega \sim J^2/4W$ due to second-neighbor resonances.  For frequencies above this shoulder, the dominant processes are these very short range hops and
the power law fit does not work, not surprisingly.  This can be seen by looking at the distribution of dipole moment matrix elements,
Fig.~\ref{dipoles}, of transitions contributing to $\sigma$: above the shoulder in (c), a peak appears at about unity, corresponding to
second-neighbor resonances, which is marked by the bottom horizontal arrow.  Third-neighbor resonances can also be
seen as a feature near $(3/2)^2=2.25$,
indicated by the upper horizontal arrow.  Data at lower values of $W$ show more such peaks, but these ``shoulders'' become less and less pronounced until they disappear altogether for $W \alt 8$.

\end{widetext}


\begin{thebibliography}{99}
\bibitem{footnote}
There are also random-matrix theory tails---i.e., short-distance resonances with anomalously small matrix element---whose contribution is subleading at low $\omega$.
\bibitem{inverse}
Because of where $\zeta$ appears in the exponentials, the proper average of $\zeta$ over a rare region is the inverse of the average of $1/\zeta$.
\end{thebibliography}
\end{document}